\documentclass[twocolumn,floatfix,superscriptaddress,nofootinbib]{revtex4}
\usepackage{graphicx}
\usepackage{amsmath}
\usepackage{amssymb}
\usepackage{bm}

\begin{document}

\title{Effect of a tunnel barrier on the scattering from a Majorana bound state in an Andreev billiard}
\author{M. Marciani}
\affiliation{Instituut-Lorentz, Universiteit Leiden, P.O. Box 9506, 2300 RA Leiden, The Netherlands}
\author{H. Schomerus}
\affiliation{Department of Physics, Lancaster University, LA1 4YB Lancaster, United Kingdom}
\author{C. W. J. Beenakker}
\affiliation{Instituut-Lorentz, Universiteit Leiden, P.O. Box 9506, 2300 RA Leiden, The Netherlands}
\date{June 2015}

\begin{abstract}
We calculate the joint distribution $P(S,Q)$ of the scattering matrix $S$ and time-delay matrix $Q=-i\hbar S^\dagger dS/dE$ of a chaotic quantum dot coupled by point contacts to metal electrodes. While $S$ and $Q$ are statistically independent for ballistic coupling, they become correlated for tunnel coupling. We relate the ensemble averages of $Q$ and $S$ and thereby obtain the average density of states at the Fermi level. We apply this to a calculation of the effect of a tunnel barrier on the Majorana resonance in a topological superconductor. We find that the presence of a Majorana bound state is hidden in the density of states and in the thermal conductance if even a single scattering channel has unit tunnel probability. The electrical conductance remains sensitive to the appearance of a Majorana bound state, and we calculate the variation of the average conductance through a topological phase transition.\\
{\tt Contribution for the special issue of Physica E in memory of Markus B\"{u}ttiker.}
\end{abstract}

\maketitle

\section{Introduction}
\label{intro}

The quantum states of particle and anti-particle excitations in a superconductor (Bogoliubov quasiparticles) are related by a unitary transformation, which means that they can be represented by a \textit{real} wave function. In this so-called Majorana representation the $N\times N$ scattering matrix $S$ at the Fermi level is real orthogonal rather than complex unitary \cite{Alt97}. Since the orthogonal group ${\rm O}(N)$ is doubly connected, this immediately implies a twofold distinction of scattering problems in a superconductor: The subgroup ${\rm O}_+(N)\equiv {\rm SO}(N)$ of scattering matrices with determinant $+1$, connected to the unit matrix, is called \textit{topologically trivial}, while the disconnected set ${\rm O}_-(N)$ of scattering matrices with determinant $-1$ is called \textit{topologically nontrivial}. In mathematical terms, the experimental search for Majorana bound states can be called a search for systems that have ${\rm Det}\,S=-1$. This search has been reviewed, from different perspectives, in Refs.\ \onlinecite{Ali12,Lei12,Sta13,Bee13,Das15}.

If the scattering is chaotic the scattering matrix becomes very sensitive to microscopic details, and it is useful to develop a statistical description: Rather than studying a particular $S$, one studies the probability distribution $P(S)$ in an ensemble of chaotic scatterers. This is the framework of random-matrix theory (RMT) \cite{Wig67,Meh64,For10}. The ensemble generated by drawing $S$ uniformly from the unitary group ${\rm U}(N)$, introduced by Dyson in the context of nuclear scattering \cite{Dys62}, is called the circular unitary ensemble (CUE). Superconductors need a new ensemble. A natural name would have been the circular orthogonal ensemble (COE), but since that name is already taken for the coset ${\rm U}(N)/{\rm O}(N)$, the alternative name circular real ensemble (CRE) is used when $S$ is drawn uniformly from ${\rm O}(N)$. The RMT of the CRE, and the physical applications to Majorana fermions and topological superconductors, have been reviewed recently \cite{Bee14}.

The uniformity of the distribution requires ideal coupling of the scattering channels to the continuum, which physically means that the discrete spectrum of a quantum dot is coupled to metal electrodes by ballistic point contacts. If the point contact contains a tunnel barrier, then $P(S)$ is no longer uniform but biased towards the reflection matrix $r_{\rm B}$ of the barrier. The modified distribution $P_{\rm Poisson}(S)$ is known \cite{Kri67,Mel85,Fri85,Bro95,Ber09}, it goes by the name ``Poisson kernel'' and equals
\begin{equation}
P_{\rm Poisson}(S)\propto {\rm Det}\,(1-r_{\rm B}^{\dagger}S)^{1-N}\label{PPoissonCRE}
\end{equation}
in the CRE \cite{Ber09}.

\begin{figure}[tb]
\centerline{\includegraphics[width=0.9\linewidth]{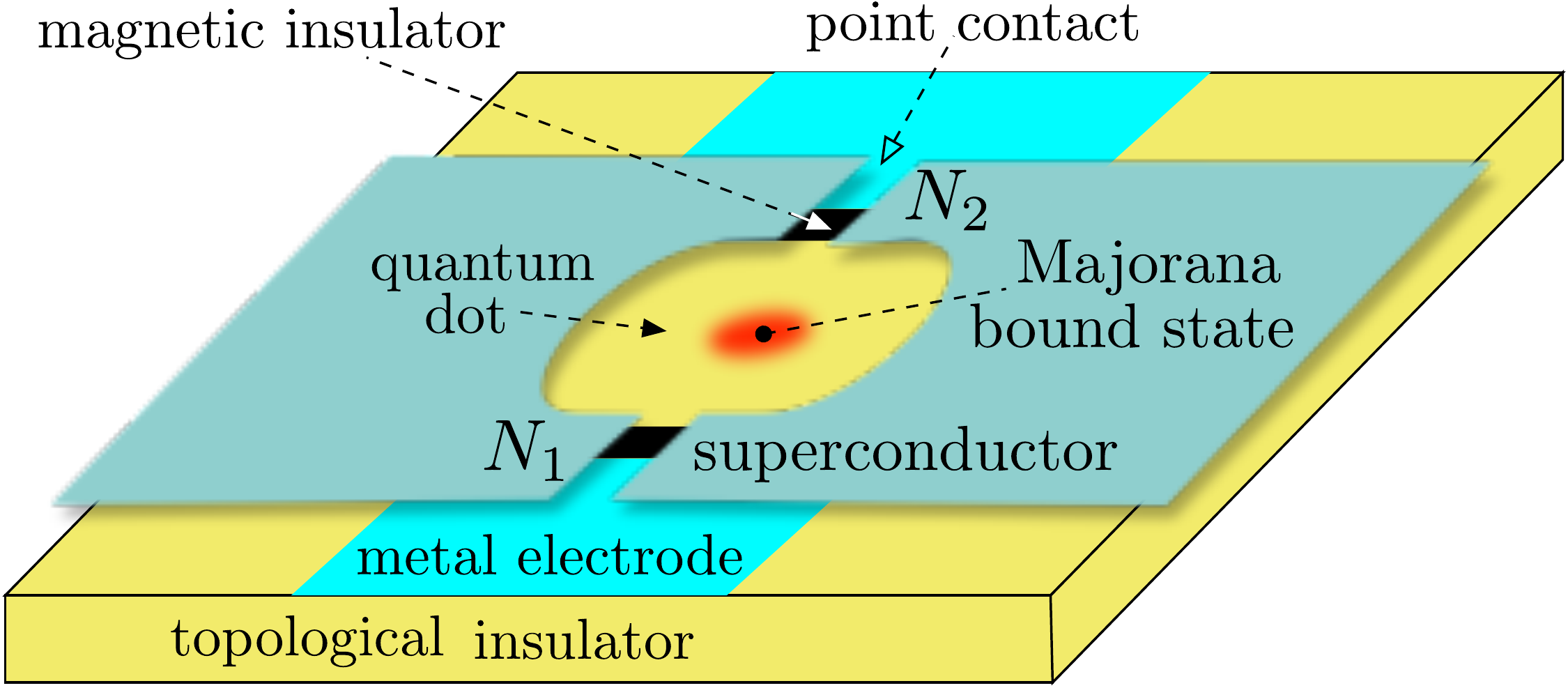}}
\caption{Andreev billiard on the conducting surface of a three-dimensional topological insulator. The billiard consists of a confined region (quantum dot, mean level spacing $\delta_0$) with superconducting boundaries, connected to metal electrodes by a pair of point contacts (supporting a total of $N=N_1+N_2$ propagating modes). A magnetic insulator introduces a tunnel barrier in each point contact (transmission probability $\Gamma$ per mode). A magnetic vortex may introduce a Majorana bound state in the quantum dot.
}
\label{fig_layout}
\end{figure}

In the present work we apply this result to the scattering (Andreev reflection) in a superconducting quantum dot (Andreev billiard), see Fig.\ \ref{fig_layout}. We focus in particular on the effect of a bound state at the Fermi level ($E=0$) in the quantum dot, a so-called Majorana zero-mode or Majorana bound state. In addition to the scattering matrix, which determines the thermal and electrical conductance, we consider also the time-delay matrix $Q=-i\hbar S^\dagger dS/dE$. The eigenvalues of $Q$ are positive numbers with the dimension of time, that govern the low-frequency dynamics of the system (admittance and charge relaxation \cite{Gop96,Bro97a,But02}). Moreover, the trace of $Q$ gives the density of states and $Q$ and $S$ together determine the thermopower \cite{Bro97b,God99}.

The joint distribution of $S$ and $Q$ is known for ballistic coupling \cite{Bro97,Mar14,Sch15}, here we generalize that to tunnel coupling. The effect of a tunnel barrier on the time-delay matrix has been studied for complex scattering matrices \cite{Sav01,Som01}, but not yet for real matrices. One essential distinction is that the tunnel barrier has no effect on the density of states in the CUE and COE, but it does in the CRE.

The outline of the paper is as follows. The next two sections formulate the scattering theory of the Andreev billiard and the appropriate random-matrix theory. Our key technical result, the joint distribution $P(S,Q)$, is given in Sec.\ \ref{sect_PS0Q}. We apply this to the simplest single-channel case ($N=1$) in Sec.\ \ref{tauDOS}, where obtain a remarkable scaling relation: For a high tunnel barrier (transmission probability $\Gamma\ll 1$) the distribution $P(\rho|\Gamma)$ of the density of states at the Fermi level is described by a one-parameter scaling function $F(x)$:
\begin{equation}
P(\rho|\Gamma)\propto \begin{cases}
F(\Gamma\rho/4)&\text{with a Majorana bound state,}\\
F(4\rho/\Gamma)&\text{without a Majorana.}
\end{cases}\label{PrhoGamma}
\end{equation}

The average density of states in the multi-channel case is calculated in Sec.\ \ref{sec_DOS}. By relating the ensemble averages of $Q$ and $S$ we derive the relation
\begin{equation}
\langle\rho\rangle=\langle\rho\rangle_{\rm ballistic}\,\left(1-\frac{2}{N\Gamma}{\rm Tr}\,r_{\rm B}^\dagger[\langle S\rangle-r_{\rm B}]\right),\label{rhoaveragegeneral}
\end{equation}
for a mode-independent tunnel probability $\Gamma$. In the CUE and COE the average scattering matrix $\langle S\rangle$ is just equal to $r_{\rm B}$, so $\langle\rho\rangle$ remains equal to its ballistic value $\langle\rho\rangle_{\rm ballistic}$, but the CRE is not so constrained. 

Applications to the thermal conductance $g$ and the electrical (Andreev) conductance $g_{\rm A}$ follow in Secs.\ \ref{transmissionG} and \ref{electrcond}. For ballistic coupling it is known that $P(g)$ is the same with or without the Majorana bound state \cite{Dah10}. (This also holds for $P(\rho)$ \cite{Mar14}.) In the presence of a tunnel barrier this is no longer the case, but we find that the Majorana bound state remains hidden if even a single scattering channel has $\Gamma=1$. The distribution of $g_{\rm A}$, in contrast, is sensitive to the presence or absence of the Majorana bound state even for ballistic coupling \cite{Bee11}. The way in which $P(g_{\rm A})$ changes as we tune the system through a topological phase transition, at which a Majorana bound state emerges, is calculated in Sec.\ \ref{phasetransition}. We conclude in Sec.\ \ref{conclude}.

In the main text we focus on the results and applications. Details of the calculations are moved to the Appendices. These also contain more general results for other RMT ensembles, with or without time-reversal and/or spin-rotation symmetry. (Both symmetries are broken in the CRE.)

\section{Scattering formulation}
\label{scattering}

Fig.\ \ref{fig_layout} shows the scattering geometry, consisting of a superconducting quantum dot (Andreev billiard) on the surface of a topological insulator, connected to normal metal electrodes by point contacts. The Hamiltonian $H$ of the quantum dot is related to the energy-dependent scattering matrix $S(E)$ by the Mahaux-Weidenm\"{u}ller formula \cite{Mah68},
\begin{equation}
\begin{split}
S(E)&=\frac{1-i\pi W^{\dagger}(E-H)^{-1}W}{1+i\pi W^{\dagger}(E-H)^{-1}W}\\
&=1-2\pi iW^{\dagger}(E-H+i\pi WW^{\dagger})^{-1}W.
\end{split}\label{SHeq}
\end{equation}
The $M\times N$ matrix $W$ couples the $M$ energy levels in the quantum dot (mean level spacing $\delta_0$) to a total of $N\ll M$ propagating modes in the point contact.

We assume that degeneracies are broken by spin-orbit coupling in the topological insulator in combination with a magnetic field (perpendicular to the surface). All degrees of freedom are therefore counted separately in $N$ and $M$, as well as in $\delta_0$. The electron-hole degree of freedom is also included in the count, but we leave open the possibility of an unpaired Majorana fermion --- a coherent superposition of electron and hole quasiparticles that does not come with a distinct antiparticle. An odd level number $M$ indicates the presence of a Majorana bound state in the quantum dot, produced when a magnetic vortex enters \cite{Fu08}. An odd mode number $N$ signals a propagating Majorana mode in the point contact, allowed by a $\pi$-phase difference between the superconducting boundaries \cite{Tit07}.

The $N$ modes have an energy-independent transmission probability $\Gamma_n\in[0,1]$ per mode. If we choose a basis such that the coupling matrix $W$ has only nonzero elements on the diagonal, it has the explicit form \cite{Ver85}
\begin{equation}
\begin{split}
&W_{mn}=w_n\delta_{mn},\;\;1\leq m\leq M,\;\;1\leq n\leq N,\\
&|w_n|^2=\frac{M\delta_0\kappa_n}{\pi^2},\;\;\kappa_n=\frac{1-r_n}{1+r_n},\;\;r_n^2=1-\Gamma_n.
\end{split}
\label{Wnmdef}
\end{equation}

Notice that the tunnel probability $\Gamma_n$ determines the reflection amplitude $r_n\in[-1,1]$ up to a sign. The conventional choice is to take $r_n\geq 0$, when $\kappa_n=\kappa_n^+$ can be written as
\begin{equation}
\kappa_n^+=\frac{1}{\Gamma_n}(2-\Gamma_n-2\sqrt{1-\Gamma_n}).\label{kappaplus}
\end{equation}
Alternatively, if $r_n\leq 0$ one has $\kappa_n=\kappa_n^-$ given by
\begin{equation}
\kappa_n^-=\frac{1}{\Gamma_n}(2-\Gamma_n+2\sqrt{1-\Gamma_n})=1/\kappa_n^+.\label{kappamin}
\end{equation}
The two choices are equivalent for ballistic coupling, $\Gamma_n=1=\kappa_n^\pm$, but for a high tunnel barrier $\Gamma_n\ll 1$ one has $\kappa_n^+\rightarrow 0$ while $\kappa_n^-\rightarrow\infty$. The sign change of $r_n$ is a topological phase transition \cite{Akh11}, which we will analyze in Section \ref{phasetransition}. For now we take $r_n\geq 0$ for all $n$, so $\kappa_n=\kappa_n^+$.

From the scattering matrix we can obtain transport properties, such as the electrical and thermal conductance, and thermodynamic properties, such as the density of states. If we restrict ourselves to properties at the Fermi level, $E=0$, we need the matrix $S(0)\equiv S$ and the derivative
\begin{equation}
Q=-i\hbar\lim_{E\rightarrow 0}S^\dagger(E)\frac{dS(E)}{dE}.\label{Qdef}
\end{equation}
The unitarity of $S(E)$ implies that $Q$ is Hermitian, so it has real eigenvalues $\tau_n$ with the dimension of time. The $\tau_n$'s are called (proper) delay times and $Q$ is called the Wigner-Smith time-delay matrix \cite{Wig55,Smi60,Fyo97}. The Fermi-level density of states $\rho$ is obtained from $Q$ via the Birman-Krein formula \cite{Bir62,Akk91,Leh95},
\begin{equation}
\rho=\frac{1}{2\pi i}\lim_{E\rightarrow 0}\frac{d}{dE}\ln{\rm Det}\,S(E)=\frac{1}{2\pi\hbar}{\rm Tr}\,Q.\label{rho0def}
\end{equation}

For the thermal conductance we partition the modes into two sets, $N=N_1+N_2$, each set connected to a different terminal, and decompose the scattering matrix into reflection and transmission subblocks,
\begin{equation}
S=\begin{pmatrix}
r&t'\\
t&r'
\end{pmatrix}.\label{S0rtdef}
\end{equation}
A small temperature difference $\delta T$ between the two terminals, at average temperature $T_0$, drives a heat current $J=G_{\rm thermal}\delta T$. The thermal conductance $G_{\rm thermal}$ in the low-temperature linear-response limit $T_0,\delta T/T_0\rightarrow 0$ is given by
\begin{equation}
g =G_{\rm thermal}/G_0={\rm Tr}\,tt^\dagger,\;\;G_0=\frac{\pi^2 k_{\rm B}^2 T_0}{6h}.\label{Gthermaldef}
\end{equation}
The quantum $G_0$ is a factor-of-two smaller than in systems without superconductivity \cite{Sch00}, due to our separate counting of electron and hole degrees of freedom that allows to account for the possibility of propagation via an unpaired Majorana mode.

If we keep the two terminals at the same temperature but instead apply a voltage difference, we can drive an electrical current. We consider a situation where both terminal 2 and the superconductor are grounded, while terminal 1 is biased at voltage $V$. The current $I$ from terminal 1 to ground is then given by the Andreev conductance
\begin{equation}
\begin{split}
g_{\rm A}=\frac{h}{e^2}\frac{dI}{dV}&={\rm Tr}\,(1-r_{ee}^{\vphantom{\dagger}}r_{ee}^\dagger+r_{he}^{\vphantom{\dagger}}r_{he}^\dagger)\\
&=\tfrac{1}{2}\,{\rm Tr}\,(1-r\tau_z r^{\dagger}\tau_z),\label{dI1dV1}
\end{split}
\end{equation}
in the zero-temperature, zero-voltage limit. In the last equality we used the particle-hole symmetry relation $t=\tau_x t^\ast\tau_x$ at $E=0$, where the $\tau_i$ Pauli matrices act on the electron ($e$) and hole ($h$) degree of freedom.

\section{Random-matrix formulation}
\label{RMT}

For a statistical description we consider an ensemble of quantum dots, each with its own random Hamiltonian $H$. The mean level spacing $\delta_0$ and coupling matrix $W$ are kept fixed. If the wave dynamics in the quantum dot is chaotic, the ensemble is fully characterized by the presence or absence of certain fundamental symmetries. This is the universal framework of random-matrix theory.

Superconducting systems are characterized by particle-hole symmetry,
\begin{equation}
\begin{split}
&H=-\tau_x H^\ast\tau_x,\;\;W=\tau_x W^\ast\tau_x,\\
&\Rightarrow S=\tau_x S^\ast\tau_x,\;\;Q=\tau_x Q^\ast\tau_x.
\end{split}
\label{ehsymmetry}
\end{equation}
The Pauli matrices $\tau_x$ can be removed from the symmetry relation by a unitary transformation
\begin{equation}
H\mapsto \Omega H\Omega^{\dagger},\;\;\Omega=\sqrt{\tfrac{1}{2}}\begin{pmatrix}
1&1\\
i&-i
\end{pmatrix},\label{SOmega}
\end{equation}
after which we simply have
\begin{equation}
H=-H^\ast,\;\; W=W^\ast,\;\; S=S^\ast,\;\; Q=Q^\ast.\label{symmetryMajoranabasis} 
\end{equation}
In this so-called Majorana basis the Hamiltonian is real antisymmetric, $H=iA$ with $A_{nm}=A_{nm}^\ast=-A_{mn}$.

If no other symmetries are imposed on the Hamiltonian we have the class-D ensemble of random-matrix theory \cite{Alt97,Bee14}, with Gaussian probability distribution
\begin{equation}
P(\{A_{nm}\})\propto\prod_{n>m}\exp\left(-\frac{\pi^{2}A_{nm}^{2}}{2M\delta_0^{2}}\right).\label{GaussEns}
\end{equation}
The eigenvalues of the antisymmetric $M\times M$ matrix $H$ come in $\pm E$ pairs, hence if $M$ is odd there must be a nondegenerate eigenvalue $E=0$ at the Fermi level, in the middle of the superconducting gap. This so-called Majorana bound state is the hallmark of a topologically nontrivial superconductor \cite{Boc00,Iva02}. If $M$ is even there is no level pinned to $E=0$, and the superconductor is called topologically trivial. It is helpful to encode the distinction in a topological quantum number $\nu$ that counts the number of Majorana bound states, so $\nu$ equals 0 or 1 if the superconductor is topologically trivial or non-trivial, respectively.

In the scattering matrix the presence of a Majorana bound state is signaled by the sign of the determinant, 
\begin{equation}
{\rm Det}\,S=(-1)^\nu,\;\;\nu\in\{0,1\}.\label{DetS0nurelation}
\end{equation} 
This can be seen directly from the definition \eqref{SHeq} in the Majorana basis: For $M$ even the matrix $H=iA$ is invertible, so we have
\begin{equation}
S=\frac{1+\pi W^{\rm T}A^{-1}W}{1-\pi W^{\rm T}A^{-1}W}\Rightarrow{\rm Det}\,S=+1,\label{S0Meven}
\end{equation}
since ${\rm Det}\,(1+{\cal A})={\rm Det}\,(1-{\cal A})$ if ${\cal A}=-{\cal A}^{\rm T}$. For $M$ odd the bound state contributes to the determinant a factor
\[
\lim_{\epsilon\rightarrow 0}\frac{{\rm Det}\,(1+\epsilon^{-1}vv^{\rm T})}{{\rm Det}\,(1-\epsilon^{-1}vv^{\rm T})}=-1,
\]
for some vector $v$, so ${\rm Det}\,S=-1$.

The class-D ensemble of scattering matrices thus consists of two disjunct sets: The special orthogonal group ${\rm SO}(N)\equiv {\rm O}_{+}(N)$ of orthogonal matrices with determinant $+1$ in the topologically trivial case, and the complement ${\rm O}_-(N)$ of orthogonal matrices with determinant $-1$ in the topologically nontrivial case.

\section{Joint distribution of scattering matrix and time-delay matrix}
\label{sect_PS0Q}

For ballistic coupling ($\Gamma_n=1$ for all $n$) the matrices $S$ and $Q$ are statistically independent \cite{Bro97}, so they can be considered separately. The class-D ballistic scattering matrix is uniformly distributed in ${\rm O}_\pm(N)$ --- uniformity being defined with respect to the Haar measure \cite{Alt97,Bee14}. This is the Circular Real Ensemble (CRE), the analogue for real orthogonal matrices of the Circular Unitary Ensemble (CUE) for complex unitary matrices \cite{Meh64,For10,Dys62}. 

The class-D ballistic time-delay matrix has probability distribution \cite{Mar14},
\begin{equation}
P_{\rm ballistic}(Q)\propto({\rm Det}\,Q)^{-3N/2}\Theta(Q)\exp(-\tfrac{1}{2}\tau_{\rm  H}\,{\rm Tr}\,Q^{-1}),\label{PQballistic}
\end{equation}
where $t_{\rm H}=2\pi\hbar/\delta_0$ is the Heisenberg time and $\Theta(Q)$ restricts $Q$ to positive definite real symmetric matrices. This constraint can be implemented more directly by defining
\begin{equation}
Q^{-1}=t_{\rm H}^{-1}\,KK^{\rm T},\;\;K\in\mathbb{R}_{N,2N-1},\label{QWrelation}
\end{equation}
with $\mathbb{R}_{n,m}$ the set of $n\times m$ matrices with real elements. The distribution \eqref{PQballistic} is then equivalent\footnote{
To transform from $P(Q)$ to $P(K)$ multiply by the Jacobians $||\partial Q/\partial Q^{-1}||\times||\partial KK^{\rm T}/\partial K||=({\rm Det}\,Q)^{N+1}\times({\rm Det}\,K)^{2-N}\propto({\rm Det}\,Q)^{3N/2}$, for $Q=Q^{\rm T}\in\mathbb{R}_{N,N}$ and $K\in\mathbb{R}_{N,2N-1}$.}
to the Wishart distribution \cite{For10}
\begin{equation}
P_{\rm Wishart}(K)\propto\exp(-\tfrac{1}{2}\,{\rm Tr}\,KK^{\rm T}).\label{ballisticWishart}
\end{equation}
Remarkably, there is no dependence on the topological quantum number for ballistic coupling: $Q$ has the same distribution irrespective of the presence or absence of a Majorana bound state.

Tunnel coupling is described by a reflection matrix $r_{\rm B}$ (from outside to outside) and transmission matrix $t_{\rm B}$ (from outside to inside). In the Majorana basis these are real matrices, parameterized by
\begin{equation}
\begin{split}
&r_{\rm B}=O_1\,{\rm diag}\,(r_1,r_2,\ldots r_N)O_2,\\
&t_{\rm B}=O_3\,{\rm diag}\,(\Gamma_1^{1/2},\Gamma_2^{1/2},\ldots \Gamma_N^{1/2})O_2,\\
&O_1,O_2,O_3\in{\rm SO}(N),\;\;\Gamma_n=1-r_{n}^2\in(0,1].
\end{split}
\label{rBdef}
\end{equation}
As we derive in App.\ \ref{app_PS0Q}, the matrix product
\begin{equation}
{\Sigma}=(1-S^{\rm T} r_{\rm B})t_{\rm B}^{-1}\label{Mdef}
\end{equation}
determines the joint distribution
\begin{subequations}
\label{PS0Q}
\begin{align}
P(S,Q)\propto{}&({\rm Det}\,{\Sigma})^{N}\,({\rm Det}\,Q)^{-3N/2}\,\Theta(Q)\nonumber\\
&\times\exp(-\tfrac{1}{2}\tau_{\rm  H}\,{\rm Tr}\,{\Sigma}^{\rm T}Q^{-1}{\Sigma}),\label{PS0Qa}\\
\Leftrightarrow
P(S,K)\propto{}&({\rm Det}\,{\Sigma})^{N}\exp(-\tfrac{1}{2}\,{\rm Tr}\,{\Sigma}^{\rm T} KK^{\rm T}{\Sigma}).
\label{PS0Qb}
\end{align}
\end{subequations}

As a check, we can integrate out the time-delay matrix to obtain the marginal distribution of the scattering matrix,
\begin{equation}
P(S)=\int dK\, P(S,K)\propto ({\rm Det}\,{\Sigma})^{N}\left|\left|\frac{\partial {\Sigma}^{\rm T}K}{\partial K}\right|\right|^{-1}.\label{PS0Jacobian}
\end{equation}
The Jacobian evaluates to \cite{Mat97}
\begin{equation}
\left|\left|\frac{\partial {\Sigma}^{\rm T}K}{\partial K}\right|\right|=({\rm Det}\,{\Sigma})^{2N-1}\;\;\text{for}\;\;K\in\mathbb{R}_{N,2N-1},\label{dMWdW}
\end{equation}
and we recover the class-D Poisson kernel\footnote{In most expressions for the probability distribution we write $\propto$ to indicate an unspecified normalization constant. The Poisson kernel \eqref{PRPoisson} is normalized, $\int P_{\rm Poisson}(S)\,dS=\int dS\equiv 1$ with $dS$ the Haar measure on ${\rm O}_{\pm}(N)$, so we use $=$ instead of $\propto$.}  
\cite{Ber09},
\begin{equation}
P_{\rm Poisson}(S)=({\rm Det}\,{\Sigma})^{1-N}
=\left(\frac{\prod_n\sqrt{\Gamma_n}}{{\rm Det}\,(1-r_{\rm B}^{\rm T}S)}\right)^{N-1}.\label{PRPoisson}
\end{equation}

The joint distribution \eqref{PS0Q} tells us that $S$ and $Q$ become correlated in the presence of a tunnel barrier. However, $S$ remains independent of the matrix product
\begin{equation}
Q_0=\frac{1}{{\Sigma}}Q\frac{1}{{\Sigma}^{\rm T}},\label{Q0def}
\end{equation}
so that the joint distribution of $S$ and $Q_0$ factorizes,
\begin{equation}
P(S,Q_0)=P_{\rm Poisson}(S)\times P_{\rm ballistic}(Q_0).\label{PSQ0def}
\end{equation}
The transformation from $Q$ to $Q_0$ removes the effect of the tunnel barrier on the time-delay matrix (see App.\ \ref{app_PS0Q}).

\section{Single-channel delay-time statistics}
\label{tauDOS}

For ballistic coupling the distribution \eqref{PQballistic} implies that the eigenvalues $\gamma_n\equiv 1/\tau_n$ of $Q^{-1}$ have the $\nu$-independent distibution \cite{Mar14}
\begin{equation}
\begin{split}
P_{\rm ballistic}(\{\gamma_n\})\propto\prod_{k=1}^{N}\gamma_k^{-1+N/2}\exp(-\tfrac{1}{2} t_{\rm H}\gamma_k)\theta(\gamma_k)\\
\qquad\times\prod_{i<j}|\gamma_i-\gamma_j|,\;\;\nu\in\{0,1\}.
\end{split}\label{Pballistic}
\end{equation}
The unit step function $\theta(x)$ ensures that $\gamma_n>0$ for all $n=1,2,\ldots N$.
 
In the single-channel case $N=1$ we can use the joint distribution \eqref{PS0Q} to immediately extend this result to arbitrary tunnel probability $ \Gamma=1-r_{\rm B}^2$. The scalar $S$ is pinned to $(-1)^\nu$, hence
\begin{align}
&{\Sigma}=\frac{\bigl(1-(-1)^\nu r_{\rm B}\bigr)}{\sqrt{1-r_{\rm B}^2}}=\begin{cases}
\sqrt{\kappa}&{\rm for}\;\;\nu=0,\\
1/\sqrt{\kappa}&{\rm for}\;\;\nu=1,
\end{cases}\label{MforNis1}\\
&\kappa=\frac{1}{\Gamma}(2-\Gamma-2\sqrt{1-\Gamma}).\label{kappaNis1}
\end{align}
[This definition of $\kappa$ corresponds to $\kappa^+$ from Eq.\ \eqref{kappaplus}.]

\begin{figure}[tb]
\centerline{\includegraphics[width=0.8\linewidth]{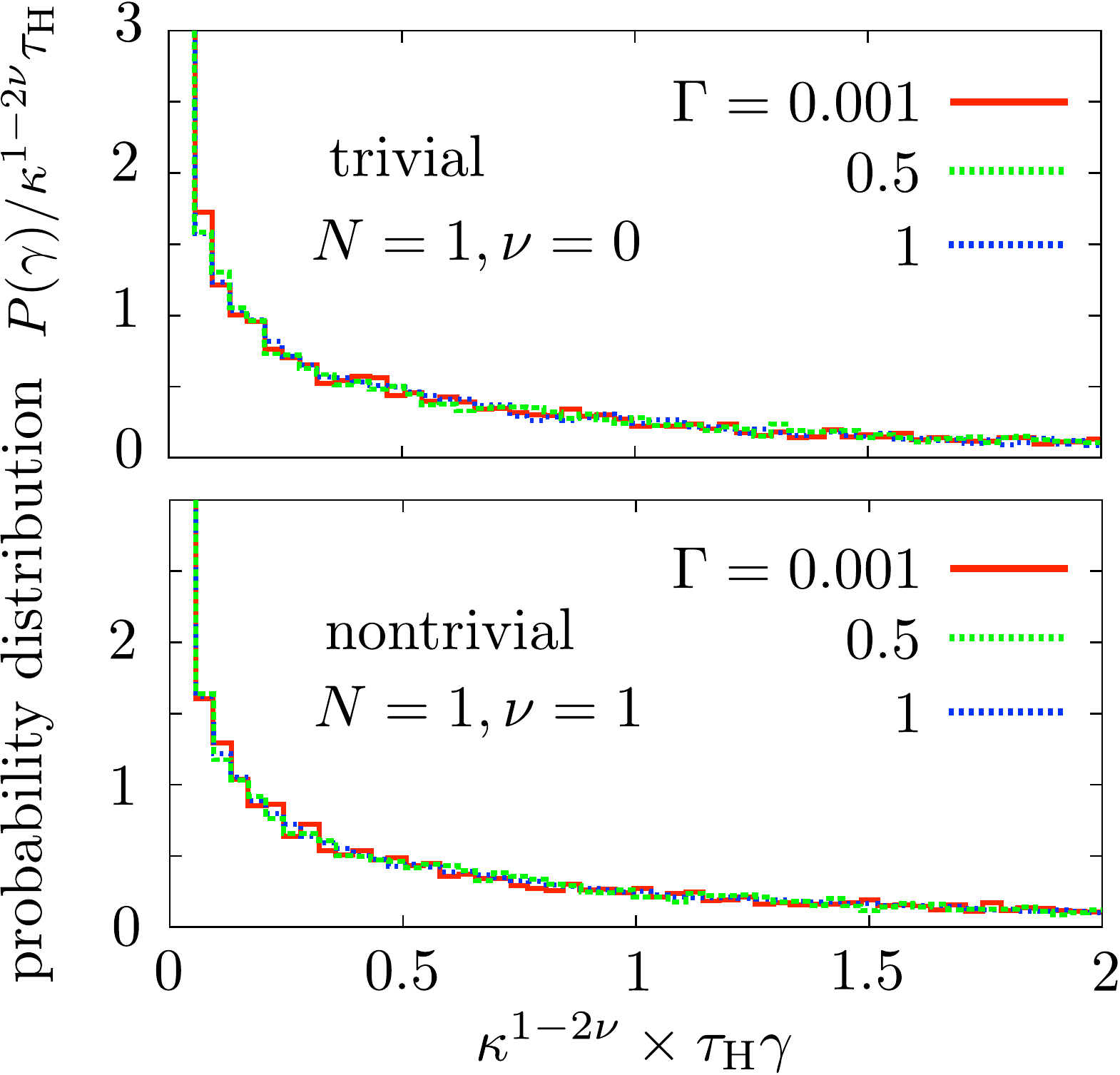}}
\caption{Probability distribution of the inverse delay time $\gamma$ for a single-channel chaotic scatterer, without ($\nu=0$) or with ($\nu=1$) a Majorana bound state. The histograms are numerical results obtained by generating random Hamiltonians (of size $M=40+\nu$) with distribution \eqref{GaussEns}. The scattering matrix, and hence the delay time, then follows from Eqs.\ \eqref{SHeq} and \eqref{Wnmdef}. The different curves correspond to different transmission probability $\Gamma$ of the tunnel barrier. Rescaling with a factor $\kappa^{1-2\nu}$, with $\kappa$ defined in Eq.\ \eqref{kappaNis1}, makes all histograms collapse onto a single curve, in agreement with the analytical result \eqref{PNis1}.
}
\label{fig_singlechannel}
\end{figure}

Since $\kappa$ then appears only as a scale factor, we conclude that the single eigenvalue $\gamma_1\equiv\gamma$ of $Q^{-1}$ for $N=1$ and any $\kappa\in(0,1]$ has distribution
\begin{equation}
P(\gamma)=\frac{\theta(\gamma)t_{\rm H}}{\sqrt{2\pi t_{\rm H}\gamma}}\times\begin{cases}
\kappa^{1/2}\exp(-\tfrac{1}{2}\kappa \,t_{\rm H}\gamma)&{\rm for}\;\;\nu=0,\\
\kappa^{-1/2}\exp(-\tfrac{1}{2}\kappa^{-1} t_{\rm H}\gamma)&{\rm for}\;\;\nu=1.
\end{cases}\label{PNis1}
\end{equation}
The single-parameter scaling $P(\gamma|\kappa,\nu)=\kappa^{1-2\nu}F(\kappa^{1-2\nu}\gamma)$ is tested numerically in Fig.\ \ref{fig_singlechannel}, by drawing random Hamiltonians from the Gaussian class-D ensemble \eqref{GaussEns}. The excellent agreement serves as a check on our analytics.

With $\rho=(2\pi\hbar\gamma)^{-1}$ the distribution \eqref{PNis1} gives the scaling form \eqref{PrhoGamma} of the density of states distribution $P(\rho|\Gamma)$ from the introduction, in the tunneling regime $\Gamma\ll 1$ when $\kappa=\Gamma/4$.

\section{Average density of states}
\label{sec_DOS}

For ballistic coupling, integration of $\rho=(2\pi\hbar)^{-1}\sum_n \gamma_n^{-1}$ with distribution \eqref{Pballistic} gives the average density of states at the Fermi level \cite{Mar14},
\begin{equation}
\langle\rho\rangle_{\rm ballistic}=\delta_0^{-1}\frac{N}{N-2},\;\;\delta_0=2\pi\hbar/\tau_{\rm H},\label{rhoballistic}
\end{equation}
for $N\geq 3$. The ensemble average diverges for $N=1,2$.

To calculate the effect of a tunnel barrier we write
\begin{equation}
\rho=(2\pi\hbar)^{-1}\,{\rm Tr}\,({\Sigma}Q_0 {\Sigma}^{\rm T}),\label{rhoQ0M}
\end{equation}
see Eq.\ \eqref{Q0def}, and then use the fact that $Q_0$ is independent of $S$ and hence independent of ${\Sigma}$. The average of $Q_0$ with distribution $P_{\rm ballistic}(Q_0)$ is proportional to the unit matrix,
\begin{equation}
\langle Q_0\rangle=\openone\,\frac{2\pi\hbar}{N} \langle\rho\rangle_{\rm ballistic}=\openone\,\frac{\tau_{\rm H}}{N-2},\label{Q0average}
\end{equation}
so the average density of states (still for $N\geq 3$) is given by
\begin{align}
\delta_0\langle\rho\rangle&=\frac{1}{N-2}\,{\rm Tr}\,\langle {\Sigma}{\Sigma}^{\rm T}\rangle\nonumber\\
&=\frac{1}{N-2}\left(\sum_n\frac{2-\Gamma_n}{\Gamma_n}-2\,{\rm Tr}\,\bigl[(t_{\rm B}^{\rm T}t_{\rm B})^{-1}r_{\rm B}^{\rm T}\langle S\rangle \bigr]\right).\label{rhoaverage}
\end{align}
(This is Eq.\ \eqref{rhoaveragegeneral} from the introduction.)

It remains to calculate the average of $S$ with the Poisson kernel \eqref{PRPoisson}. In the Wigner-Dyson symmetry classes this average is just $r_{\rm B}$, but as pointed out in Ref.\ \onlinecite{Ber09} this no longer holds in the Altland-Zirnbauer class D. A simple result for $\langle S\rangle$ is possible for mode-independent tunnel probabilities, $\Gamma_n=\Gamma$ for all $n$, see App.\ \ref{app_Saverage}:
\begin{equation}
\langle S\rangle_\pm=r_{\rm B}\left(1-N^{-1}\pm N^{-1}(1-\Gamma)^{N/2-1}\right),\label{Saverage}
\end{equation}
where again the $+$ sign corresponds to $\nu=0$ (without a Majorana bound state) and the $-$ sign to $\nu=1$ (with a Majorana bound state). 

\begin{figure}[tb]
\centerline{\includegraphics[width=0.8\linewidth]{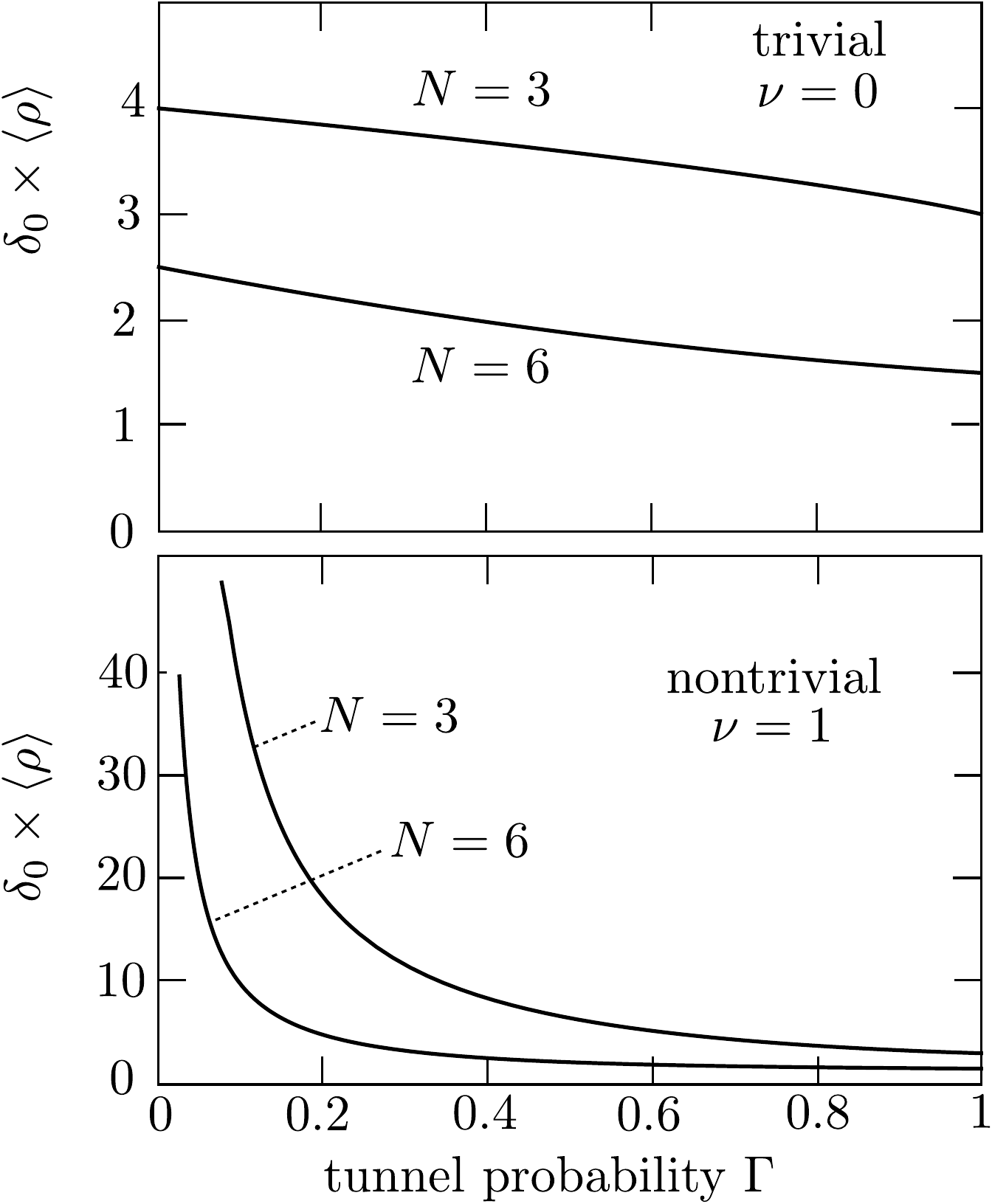}}
\caption{Ensemble averaged density of states as a function of mode-independent transmission probability through the barrier, in the absence ($\nu=0$) or in the presence ($\nu=1$) of a Majorana bound state. The curves are calculated from the analytical expression \eqref{rhoaverageresult}.
}
\label{fig_rhoaverage}
\end{figure}

We thus arrive at the average density of states,
\begin{equation}
\delta_0\langle\rho\rangle_\pm=\frac{N}{N-2}\left(1-\frac{2}{N\Gamma} \left[\Gamma-1\pm (1-\Gamma)^{N/2}\right]\right),\label{rhoaverageresult}
\end{equation}
plotted in Fig.\ \ref{fig_rhoaverage}. In the ballistic limit $\Gamma\rightarrow 1$ the dependence on the Majorana bound state drops out, while in the tunneling limit $\Gamma\rightarrow 0$ we obtain
\begin{equation}
\delta_0\langle\rho\rangle=\frac{2}{N-2}\times\begin{cases}
N-1+{\cal O}(\Gamma)&{\rm for}\;\;\nu=0,\\
2/\Gamma-1+{\cal O}(\Gamma)&{\rm for}\;\;\nu=1.
\end{cases}
\end{equation}
The $1/\Gamma$ divergence of the density of states for $\nu=1$ corresponds to the delta-function contribution from the Majorana bound state in the closed system. Without the Majorana bound state ($\nu=0$) the density of states at the Fermi level remains finite in the $\Gamma\rightarrow 0$ limit, but it does remain above the normal-state value of $1/\delta_0$. This midgap spectral peak is characteristic for a class-D superconductor \cite{Alt97,Bee14,Boc00,Iva02,Bag12}. While in a closed system the peak is simply a factor of two, in the weakly coupled open system it is a larger factor $2(N-1)/(N-2)$, which only tends to 2 in the large-$N$ limit. The fact that the ensemble of open systems does not reduce to an ensemble of closed systems in the limit $\Gamma\rightarrow 0$ is due to statistical fluctuations that remain important for small $N$.

\section{Thermal conductance}
\label{transmissionG}

We consider the thermal conductance in the simplest case $N_1=N_2=1$ of a quantum dot with single-mode point contacts. These are Majorana modes, carrying heat but no charge. 

The scattering matrix $S\in{\rm O}_\pm(2)$ is parameterized by
\begin{equation}
S_\pm=\begin{pmatrix}
\cos\theta&\mp\sin\theta\\
\sin\theta&\pm\cos\theta
\end{pmatrix},\;\;
S=\begin{cases}
S_+&{\rm if}\;\;\nu=0,\\
S_-&{\rm if}\;\;\nu=1.
\end{cases}\label{Spmdef}
\end{equation}
The Haar measure equals
\begin{equation}
d\mu=\pi^{-1}d\theta,\;\;0<\theta<\pi,\label{HaarO2}
\end{equation} 
the same for both ${\rm O}_+$ and ${\rm O}_-$. 

For the tunnel barrier we take the reflection matrix $r_{\rm B}={\rm diag}\,(r_1,r_2)$, with $r_n=\sqrt{1-\Gamma_n}\geq 0$. The Poisson kernel \eqref{PRPoisson} then has the explicit form
\begin{equation}
P_\pm(\theta)=\frac{\sqrt{\Gamma_1\Gamma_2}}{1\pm r_1 r_2-(r_1\pm r_2)\cos\theta}.\label{PSpm}
\end{equation}

The dimensionless thermal conductance \eqref{Gthermaldef} has distribution
\begin{align}
&P_\pm(g )=\frac{1}{\pi}\int_0^\pi d\theta\,\delta(g-\sin^2\theta)P_\pm(\theta)\nonumber\\
&\quad=P_{\rm ballistic}(g )\frac{(1\pm r_1 r_2)\sqrt{\Gamma_1\Gamma_2}}{(1\pm r_1 r_2)^2-(1-g)(r_1\pm r_2)^2},\label{Pgresult}
\end{align}
where as before, $P_+$ applies to $\nu=0$ and $P_-$ to $\nu=1$. The distribution
\begin{equation}
P_{\rm ballistic}(g )=\frac{1}{\pi\sqrt{g (1-g )}},\;\;0<g <1,\label{P0g}
\end{equation}
is the result \cite{Dah10} for ballistic coupling ($\Gamma_n=1$, $r_n=0$).

\begin{figure}[tb]
\centerline{\includegraphics[width=0.8\linewidth]{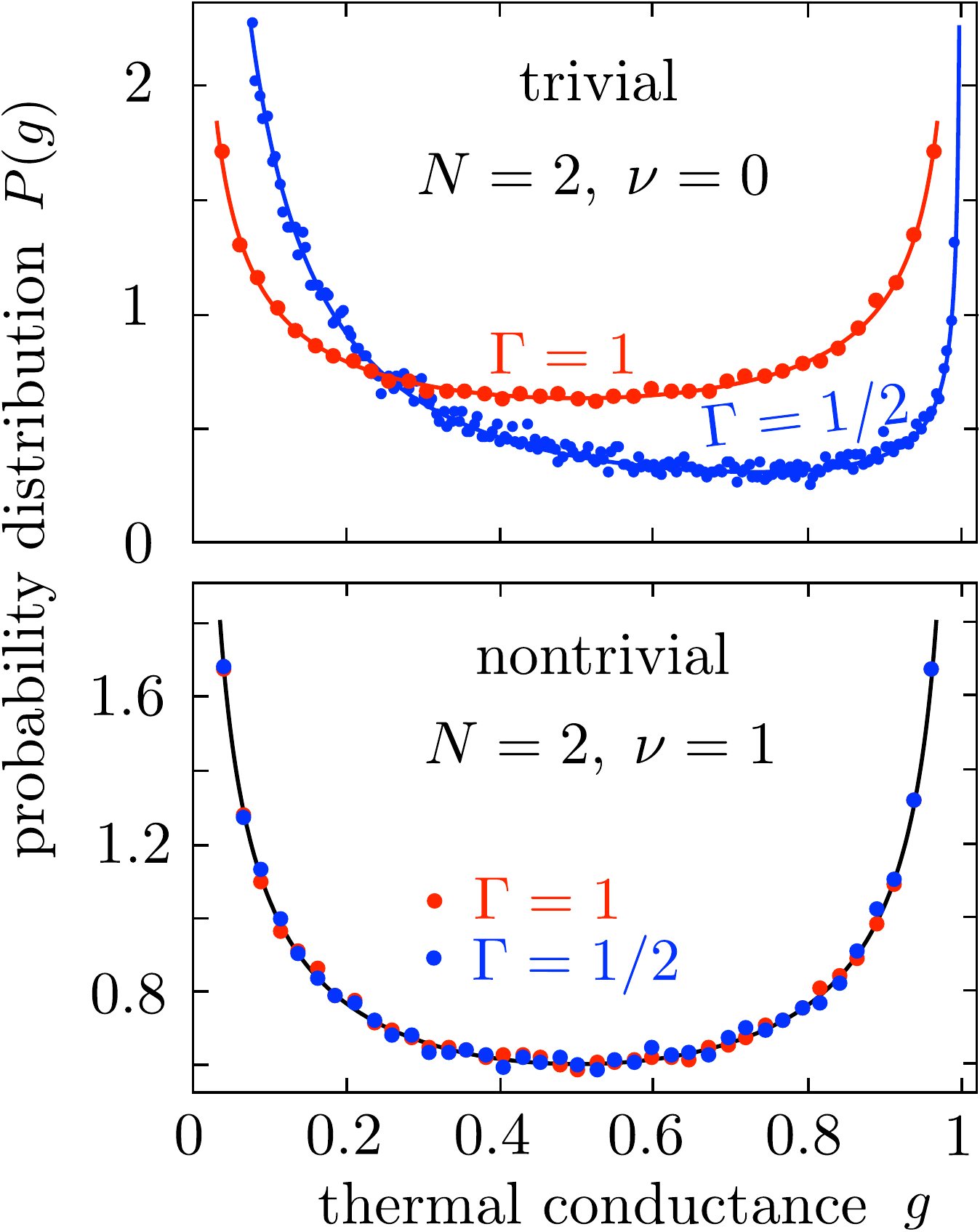}}
\caption{Probability distribution of the thermal conductance $g =G_{\rm thermal}/G_0$, see Eq.\ \eqref{Gthermaldef}, for a chaotic scatterer having two single-mode point contacts with identical tunnel probabilities $\Gamma$. The data points are numerical results for a random Hamiltonian ($M=46+\nu$), the curves are the analytical result \eqref{Pgequalbarrier}. In the presence of a Majorana zero-mode ($\nu=1$) the distribution is independent of $\Gamma$.
}
\label{fig_thermalcond}
\end{figure}

For identical tunnel barriers, $\Gamma_1=\Gamma_2=\Gamma$, this reduces to
\begin{equation}
P(g)=P_{\rm ballistic}(g)\times\begin{cases}
\frac{\Gamma(2-\Gamma)}{\Gamma^2+4g(1-\Gamma)}&{\rm if}\;\nu=0,\\
\qquad\; 1&{\rm if}\;\nu=1.
\end{cases}\label{Pgequalbarrier}
\end{equation}
Quite remarkably, the distribution of the thermal conductance for two identical single-mode point contacts is unaffected by the presence of a tunnel barrier in the topologically nontrivial case. Fig.\ \ref{fig_thermalcond} is a numerical check of this analytical result.

Notice that the distribution \eqref{Gthermaldef} becomes independent of $\nu$ if $r_1 r_2=0$. This is a special case of a more general result, valid for any $N_1,N_2$:
\begin{equation}
P_+(g)=P_-(g)\;\;{\rm if}\;\;{\rm Det}\,r_{\rm B}=0,\label{Pgtheorem}
\end{equation}
in words: The probability distribution of the thermal conductance becomes independent of the presence or absence of a Majorana bound state if the quantum dot is coupled ballistically to at least one of the scattering channels. In other words, \textit{ballistic coupling to a propagating Majorana mode hides the Majorana bound state.}
 
The proof is straightforward: If $\Gamma_{n_0}=1$ for one of the indices $n_0\in\{1,2,\ldots N\}$, then the Poisson kernel \eqref{PRPoisson} is unchanged if we multiply $S\mapsto O_1\Lambda O_1^{\rm T} S$, with $\Lambda_{nm}=\delta_{nm}(1-2\delta_{nn_0})$. [The orthogonal matrix $O_1$ is defined in Eq.\ \eqref{rBdef}.] The Haar measure remains unchanged as well, and so does the thermal conductance \eqref{Gthermaldef}. Since ${\rm Det}\,S=-{\rm Det}\,(\Lambda S)$, so ${\rm O}_+$ is mapped onto ${\rm O}_-$, we conclude that $P_+(g)=P_-(g)$.

This proof for the Poisson kernel extends to the entire joint distribution \eqref{PS0Q} of $S$ and $Q$: The transformation $S\mapsto O_1\Lambda O_1^{\rm T} S$ has no effect on the matrices $Q$ and ${\Sigma}$, so $P(S,Q)$ remains unchanged. It follows that the probability distribution of the density of states is the same with or without a Majorana bound state if $\Gamma_n=1$ for at least one of the scattering channels.

\section{Electrical conductance}
\label{electrcond}

Because a Majorana mode is charge-neutral, no electrical current can be driven for $N_1=N_2=1$. A nonzero current $I$ is possible for $N_1=2$, when terminal 1 biased at voltage $V$ has a distinct electron and hole mode. We investigate the effect of a tunnel barrier for $N_1=2$, $N_2=1$. In the Majorana basis the expression \eqref{dI1dV1} for the Andreev conductance reads
\begin{equation}
g_{\rm A}=\tfrac{1}{2}\,{\rm Tr}\,(1-r\tau_y r^{\rm T}\tau_y),\label{dI1dV1M}
\end{equation}
with $r$ a $2\times 2$ real matrix.

The scattering matrix $S_\pm\in{\rm O}_\pm(3)$ can be conveniently parameterized using three Euler angles \cite{Li12},
\begin{equation}
\begin{split}
&S_{+}=\begin{pmatrix}
R(\alpha)&0\\
0&1
\end{pmatrix}\begin{pmatrix}
1&0\\
0&R(\theta)
\end{pmatrix}\begin{pmatrix}
R(\alpha')&0\\
0&1
\end{pmatrix},\\
&S_-={\rm diag}\,(1,1,-1)\,S_+,
\end{split}
\label{SEuler}
\end{equation}
where we have defined
\begin{equation}
R(\alpha)=\begin{pmatrix}
\cos\alpha&-\sin\alpha\\
\sin\alpha&\cos\alpha
\end{pmatrix}.\label{Rthetadef}
\end{equation}
The Haar measure on ${\rm O}_\pm(3)$ is given by \cite{Mar14}
\begin{equation}
d\mu=\frac{\sin\theta}{8\pi^{2}} d\theta d\alpha d\alpha',\;\;\alpha,\alpha'\in(0,2\pi),\;\;\theta\in(0,\pi).\label{HaarO3}
\end{equation}

\begin{figure}[tb]
\centerline{\includegraphics[width=0.8\linewidth]{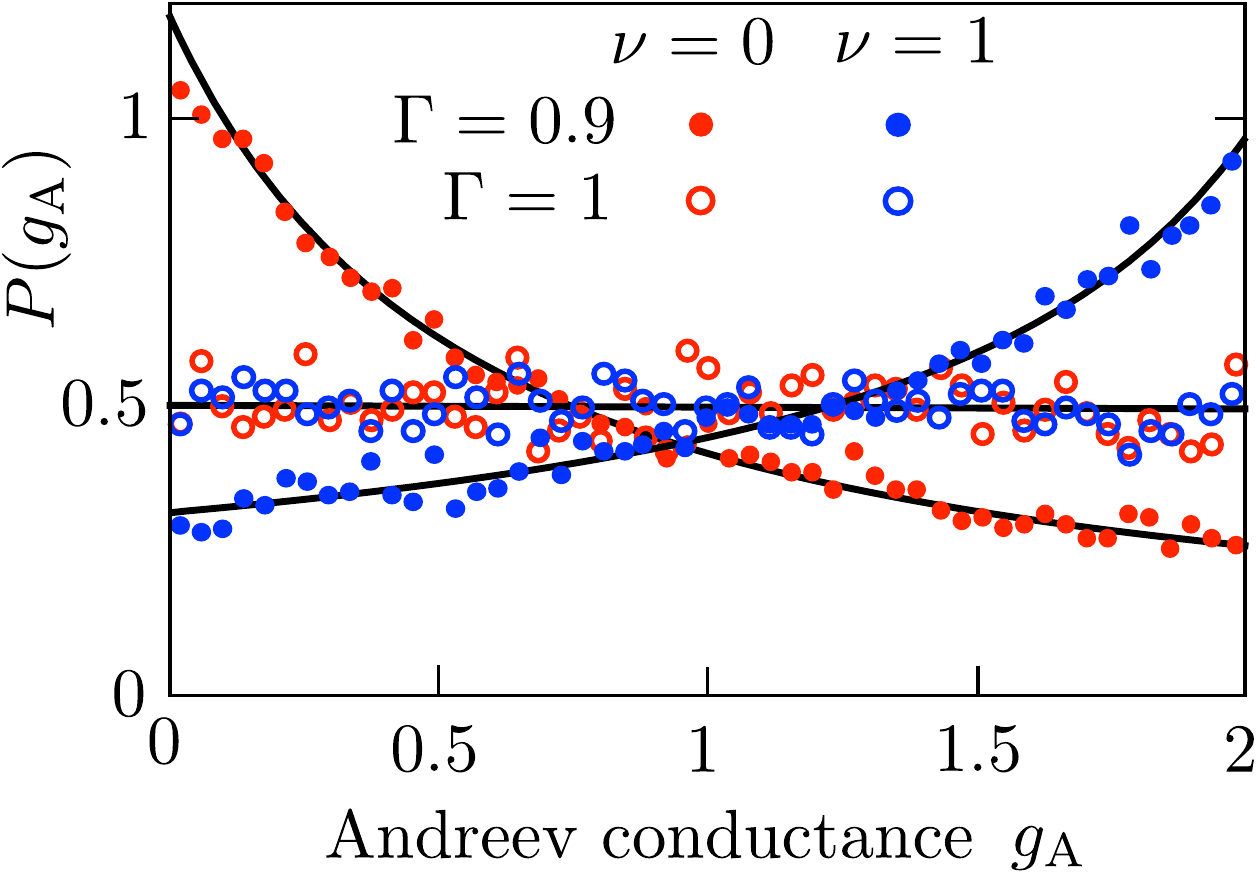}}
\caption{Probability distribution of the Andreev conductance $g_{\rm A}$, see Eq.\ \eqref{dI1dV1}, for $N_1=2$, $N_2=1$, $\Gamma_1=\Gamma_2\equiv\Gamma$. The data points are numerical results for a random Hamiltonian ($M=46+\nu$), the curves are the analytical result \eqref{Ppmg1}. For ballistic coupling ($\Gamma=1$) the distribution is the same with or without a Majorana bound state. In the limit $\Gamma\rightarrow 0$ the distribution becomes sharply peaked at $g_{\rm A}=2\nu$.
}
\label{fig_electrcond}
\end{figure}

Because $R(\alpha)$ commutes with $\tau_y$, the dimensionless conductance \eqref{dI1dV1M} depends only on the Euler angle $\theta$,
\begin{equation}
g_{\rm A}=1-\cos\theta.\label{g1O3}
\end{equation}
In point contact 1 we take a tunnel probability $\Gamma_1$, the same for the electron and hole mode, while in point contact 2 we have tunnel probability $\Gamma_2$ for the unpaired Majorana mode. Evaluation of the Poisson kernel \eqref{PRPoisson} with $r_{\rm B}={\rm diag}\,(r_1,r_1,r_2)$ gives the conductance distribution
\begin{widetext}
\begin{equation}
P_\pm(g_{\rm A})=\frac{\tfrac{1}{2}\Gamma_1^{1/2}\Gamma_2[(\Gamma_1-2)(\pm r_2-1)+(\pm r_2-1+\Gamma_1)g_{\rm A}]}{[(1\pm r_2(g_{\rm A}-1))^2-(g_{\rm A}-1\pm r_2)^2(1-\Gamma_1)]^{3/2}},\;\;0<g_{\rm A}<2,\;\;r_2=\sqrt{1-\Gamma_2},\label{Ppmg1}
\end{equation}
plotted in Fig.\ \ref{fig_electrcond} for $\nu=0$ ($P_+$), $\nu=1$ ($P_-$) and two values of $\Gamma_1=\Gamma_2\equiv \Gamma$.
\end{widetext}

In the limit $r_2\rightarrow 1$, when terminal 2 is decoupled from the quantum dot, we recover the result \cite{Ber09b}
\begin{equation}
P(g_{\rm A})=\begin{cases}
\delta(g_{\rm A})&{\rm if}\;\nu=0,\\
\delta(2-g_{\rm A})&{\rm if}\;\nu=1,
\end{cases}\label{PGamma20}
\end{equation}
independent of $\Gamma_1$. The conductance in this case is uniquely determined by the topological quantum number.

In the opposite limit $r_2\rightarrow 0$ the distribution \eqref{Ppmg1} becomes independent of $\nu$,
\begin{equation}
P(g_{\rm A})=\frac{\tfrac{1}{2}\Gamma_1^{1/2}[2-\Gamma_1-(1-\Gamma_1)g_{\rm A}]}{[1-(1-\Gamma_1)(g_{\rm A}-1)^2]^{3/2}},\;\;{\rm if}\;\;r_2=0.\label{Ppmgr20}
\end{equation}
This is a special case of a more general result, for any $N_1,N_2$, 
\begin{equation}
P_+(g_{\rm A})=P_-(g_{\rm A})\;\;{\rm if}\;\;{\rm Det}\,{\cal P}_2r_{\rm B}=0,\label{Pg1theorem}
\end{equation}
where ${\cal P}_2$ projects onto the modes coupled to terminal 2. Ballistic coupling, even for a single mode, to terminal 2 therefore removes the dependence on the topological quantum number.

The proof of Eq.\ \eqref{Pg1theorem} proceeds along the lines of the proof of Eq.\ \eqref{Pgtheorem}, with the difference that the transformation $S\mapsto O_1\Lambda O_1^{\rm T} S$ should leave the upper-left block $r$ of $S$ unaffected --- otherwise the Andreev conductance \eqref{dI1dV1M} would change. This also explains why the $\nu$-dependence of $P(g_{\rm A})$ remains for ballistic coupling to terminal 1 \cite{Bee11}.

\section{Majorana phase transition}
\label{phasetransition}

\begin{figure}[tb]
\centerline{\includegraphics[width=0.9\linewidth]{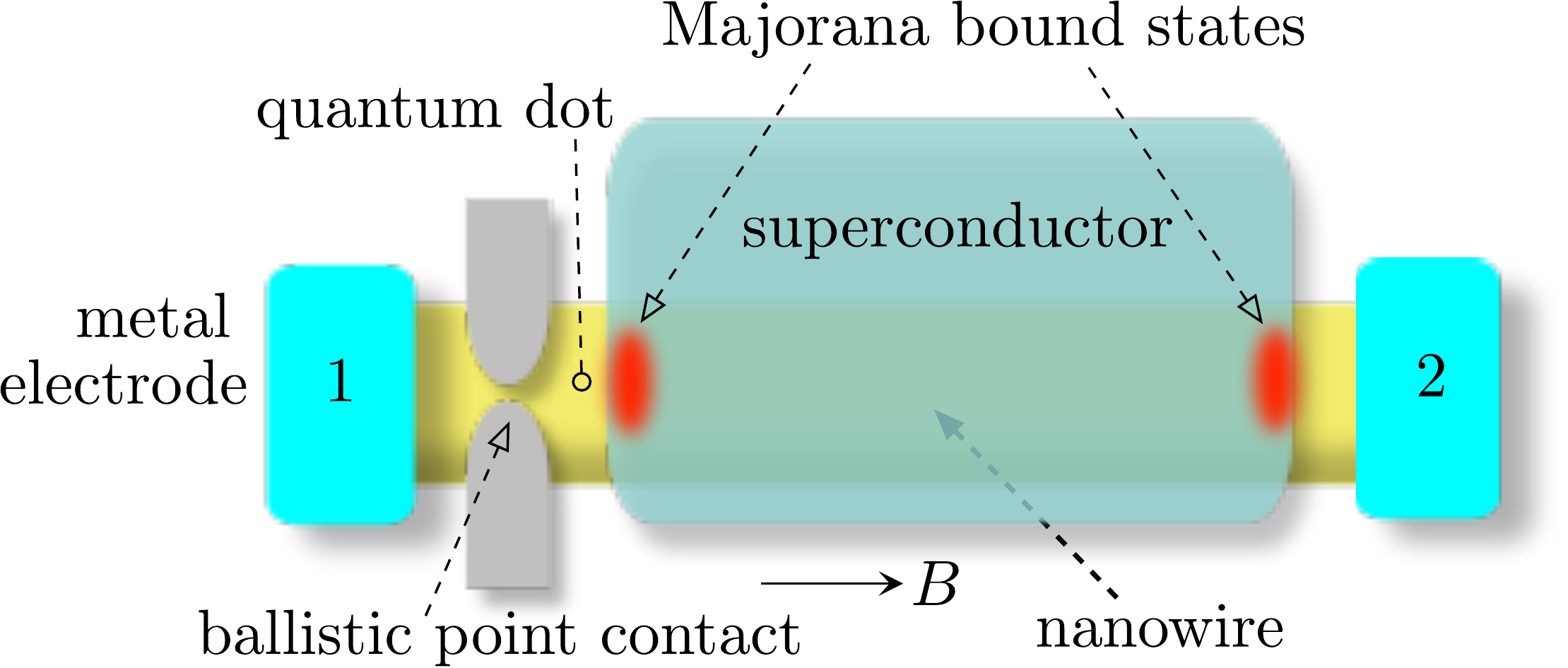}}
\caption{Device to study the topological phase transition in a semiconductor nanowire covered by a superconductor. When the Zeeman energy of a parallel magnetic field exceeds the induced superconducting gap in the nanowire, a pair of Majorana bound states emerges at the end points. One of these is coupled directly to electrode $2$, while the other is coupled to electrode $1$ via a point contact.
}
\label{fig_nanowire}
\end{figure}

The appearance of a Majorana bound state is a topological phase transition. There is a search for this transition in a nanowire geometry, see Fig.\ \ref{fig_nanowire}, where it has been predicted to occur when the Zeeman energy of a magnetic field (parallel to the wire axis) exceeds the gap induced by the proximity to a superconductor \cite{Lut10,Ore10}.

Because the Majorana bound states emerge pairwise at the two ends of the nanowire, the topological quantum number $\nu$ of the entire structure remains $0$ and the determinant ${\rm Det}\,S$ of the full scattering matrix remains $+1$ through the transition. What changes is the sign of the determinant ${\rm Det}\,r$ of the reflection submatrix. At the topological phase transition ${\rm Det}\,r=0$, implying a perfectly transmitted mode and a quantized peak in the thermal conductance \cite{Akh11}.

We study the effect of the phase transition on the statistics of the electrical conductance, measured by contacting one end of the nanowire (terminal 1) to a metal electrode at voltage $V$, while the superconductor and the other end of the nanowire (terminal 2) are at ground. Terminal 1 is connected to the nanowire via a point contact, thus creating a confined region (quantum dot) with chaotic scattering. The minimal dimensionality of the scattering matrix $S$ of the quantum dot is $3\times 3$: One electron and one hole mode connected to terminal 1 and one Majorana mode connected to terminal 2. 

To minimize the number of free parameters we assume ballistic coupling through the point contact, so the matrix $r_{\rm B}$ in the Poisson kernel \eqref{PRPoisson} is $r_{\rm B}={\rm diag}\,(1,1,r_2)$. The reflection amplitude $r_2$ at terminal 2 is tuned through zero by some external control parameter $\xi$, typically magnetic field or gate voltage. Near the transition (conveniently shifted to $\xi\equiv 0$) this dependence can be conveniently parameterized by \cite{Akh11}
\begin{equation}
r_2(\xi)=\tanh(\xi/\xi_{0}).\label{rMxi}
\end{equation}
(The width $\xi_0$ of the transition is system dependent.) The corresponding coupling constant $\kappa_2$ in Eq.\ \eqref{Wnmdef} then has an exponential $\xi$-dependence,
\begin{equation}
\kappa_2=\frac{1-r_2}{1+r_2}=\exp(-2\xi/\xi_0).\label{kappaxi}
\end{equation}

\begin{figure}[tb]
\centerline{\includegraphics[width=0.8\linewidth]{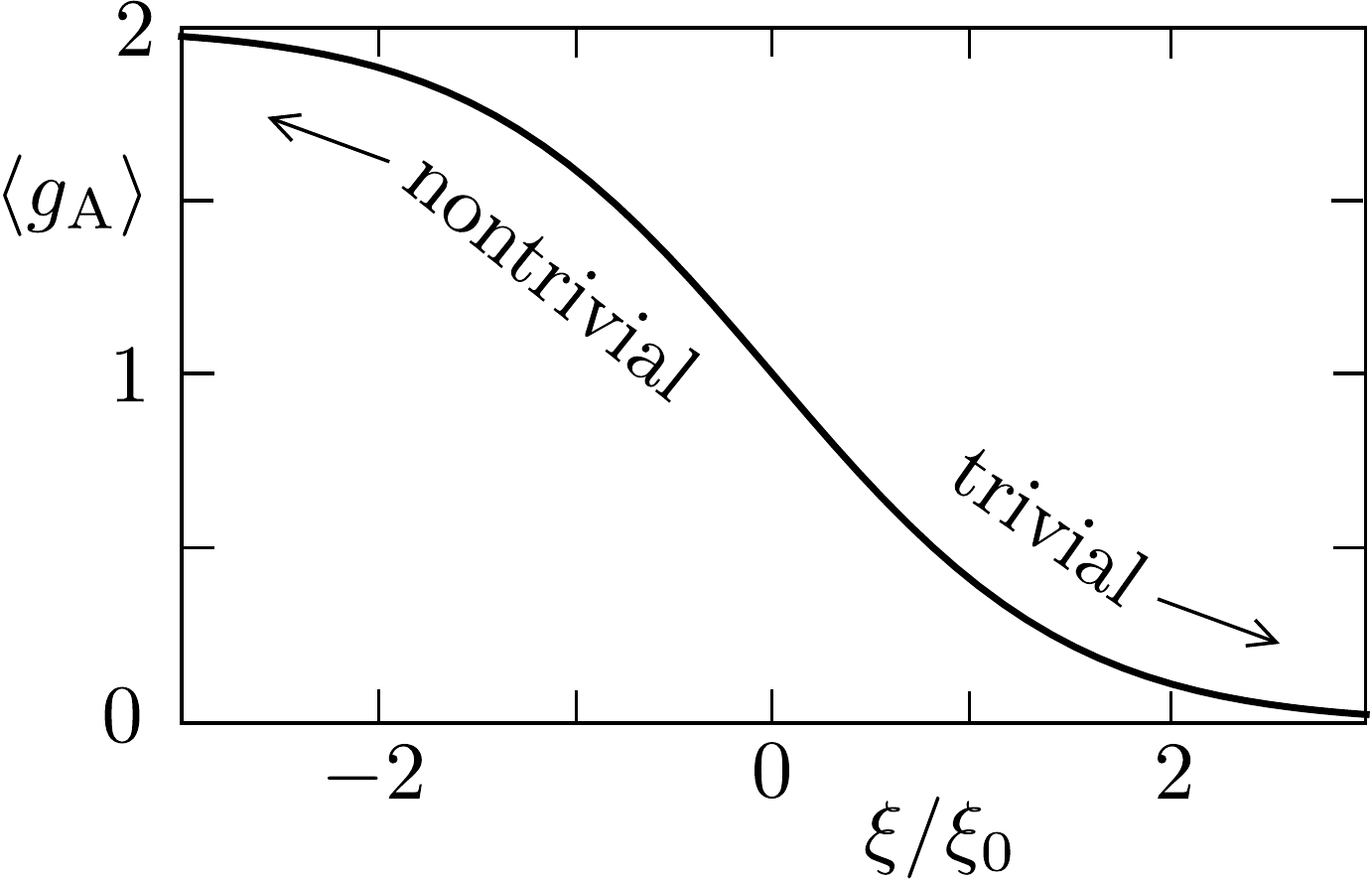}}
\caption{Variation of the ensemble-averaged Andreev conductance in the geometry of Fig.\ \ref{fig_nanowire}, as the nanowire is driven through a topological phase transition (controlled by a parameter $\xi$, which can be thought of as the deviation of the magnetic field from the critical field strength). The curve is the result \eqref{gAaverage} for a single-channel ballistic point contact.
}
\label{fig_gAtransition}
\end{figure}

The probability distribution of the Andreev conductance (in units of $e^2/h$) follows from Eq.\ \eqref{Ppmg1},
\begin{align}
P(g_{\rm A})&=\tfrac{1}{2}(1-r_2^2)[1+(g_{\rm A}-1)r_2]^{-2}\nonumber\\
&=\tfrac{1}{2}[\cosh(\xi/\xi_0)+(g_{\rm A}-1)\sinh(\xi/\xi_0)]^{-2},\nonumber\\
&\qquad\qquad\;\;0<g_{\rm A}<2.\label{Ppmg2}
\end{align}
The delta-function limits \eqref{PGamma20} are reached for $\xi\rightarrow\pm\infty$ (keeping $\nu=0$, because the entire system is topologically trivial). Right at the transition, at $\xi=0$, the distribution is uniform in the interval $0<g_{\rm A}<2$. The average conductance varies through the transition as
\begin{equation}
\langle g_{\rm A}\rangle=1-\frac{1}{\tanh(\xi/\xi_0)}+\frac{\xi/\xi_0}{\sinh^2(\xi/\xi_0)},\label{gAaverage}
\end{equation}
see Fig.\ \ref{fig_gAtransition}.

\section{Conclusion}
\label{conclude}

In conclusion, we have investigated a variety of observable consequences of the fact that the scattering matrix of Majorana fermions is real orthogonal rather than complex unitary. Of particular interest is the identification of observables that can detect the sign of the determinant, since ${\rm Det}\,S=-1$ signifies the presence of a Majorana bound state. The obvious signal of such a zero-mode, a midgap peak in the density of states \cite{Boc00,Iva02}, is broadened by tunnel coupling to the continuum. We find that the peak remains hidden if the coupling is ballistic (unit transmission) in even a single scattering channel. The thermal conductance is likewise insensitive to the presence or absence of a Majorana bound state, but the electrical conductance retains this sensitivity when the coupling is ballistic. 

These results for the effect of a tunnel barrier on the midgap spectral peak are derived from the distribution $P(S,Q)$ of scattering matrix and time-delay matrix under the assumption of chaotic scattering, due to disorder or due to irregularly shaped boundaries. The appropriate ensemble in the absence of time-reversal and spin-rotation symmetry has symmetry class D in the Altland-Zirnbauer classification \cite{Alt97}. Chiral symmetry would change this to class BDI, in which multiple zero-modes can overlap without splitting \cite{Ver85}. The effect of chiral symmetry on the joint distribution $P(S,Q)$ is known for ballistic coupling \cite{Sch15} --- but not yet for tunnel coupling. This seems a worthwhile project for future research.

\acknowledgments

This research was supported by the Foundation for Fundamental Research on Matter (FOM), the Netherlands Organization for Scientific Research (NWO/OCW), and an ERC Synergy Grant.

\appendix  

\section{Joint distribution of scattering matrix and time-delay matrix}
\label{app_PS0Q}

We calculate the joint distribution $P(S,Q)$ of scattering matrix and time-delay matrix in the presence of a tunnel barrier, starting from the known distribution $P(S_0,Q_0)$ without a barrier \cite{Bro97,Mar14}. The application in the main text concerns symmetry class D, but for the sake of generality and for later reference we give results for all four Altland-Zirnbauer \cite{Alt97} symmetry classes D, DIII,  C, CI, as well as for the three Wigner-Dyson \cite{Wig67,Dys62b} symmetry classes A, AI, AII. The symmetry indices that distinguish the ensembles are listed in Table \ref{table_AZ}, see Ref.\ \onlinecite{Bee14} for an overview of this classification.

The unitary matrix $S$ and the Hermitian matrix $Q$ are real in class D, complex in class A, and quaternion in class C. We consider these three symmetry classes without time-reversal symmetry first, and then include the constraints of time-reversal symmetry in classes DIII, CI, AI, AII.\footnote{The seven symmetry classes in Table \ref{table_AZ} do not exhaust the tenfold way classification of random-matrix theory: There are three more chiral classes \cite{Ver00} (labeled AIII, BDI, CII) that require separate consideration \cite{Sch15}.
}

\begin{table}
\centering
\begin{tabular}{ | c || c | c | c | c | c | c | c | c | c | c |}
\hline
& \multicolumn{4}{|c|}{Altland-Zirnbauer } & \multicolumn{3}{|c|}{Wigner-Dyson } \\ \hline
& \textbf{D} &  \textbf{DIII} & \textbf{C} & \textbf{CI}  & \textbf{A} & \textbf{AI} & \textbf{AII}\\ \hline \hline
$\alpha$ & $-1$  & $-1$ & $2$ & $1$  & 0 & 0 & 0  \\  \hline
$\beta$ & 1 & 2 & 4 & 2 & 2 & 1 & 4  \\ \hline
$t_0/\tau_{\rm H}$& 1 & 1 & $\tfrac{1}{2}$ & $\tfrac{1}{2}$ &1 & 1 & 1 \\ \hline
degeneracy $d$ & 1 & 2 & 2 & 2 & 1 & 1 & 2 \\ \hline
\end{tabular}
\caption{Parameters that appear in the distribution of the scattering matrix and time-delay matrix, for each of the Altland-Zirnbauer and Wigner-Dyson symmetry classes. Notice that a different set of indices $\alpha'$, $\beta'$ govern the energy level statistics \cite{Bee14}. The degeneracy factor $d$ refers to the Kramers degeneracy of the scattering channels and the delay times, ignoring uncoupled spin bands. (The energy levels may have a different degeneracy.)
}
\label{table_AZ}
\end{table}

\begin{table*}
\centering
\begin{tabular}{ | l || c | c | c | c |  c | c | c |}
\hline
& \textbf{D}  & \textbf{C} & \textbf{A} & \textbf{AI} &  \textbf{CI} & \textbf{AII} &  \textbf{DIII}  \\ \hline \hline
& \multicolumn{3}{|c|}{broken time-reversal symmetry} & \multicolumn{4}{|c|}{preserved time-reversal symmetry} \\ \hline
& \multicolumn{3}{|c|}{$\delta S\equiv S^\dagger dS=-\delta S^\dagger$} & \multicolumn{4}{|c|}{$ \delta S\equiv \tilde{U}^\dagger dS U^\dagger=-\delta S^\dagger=\delta \tilde{S}$} \\ \hline
& \multicolumn{3}{|c|}{} & \multicolumn{2}{|c|}{$\tilde{U}\equiv U^{\rm T}$} & \multicolumn{2}{|c|}{$\tilde{U}\equiv U^{\rm D}\equiv\sigma_{y}U^{\rm T}\sigma_y$} \\ \hline
$\delta S_{nm}$ & $q_0$ & $q_0\sigma_0+i\bm{q}\cdot\bm{\sigma}$   & $a+ib$ & $ia$ & $ia\sigma_x+ib\sigma_z$ & $iq_0\sigma_0+\bm{q}\cdot\bm{\sigma}$ & $a\sigma_x+b\sigma_z$  \\
 ($n\neq m$)  & $\beta=1$ & $\beta=4$  & $\beta=2$ & $\beta=1$ & $\beta=2$ & $\beta=4$ & $\beta=2$ \\ \hline
$\delta S_{nn}$  & $0$& $i\bm{q}\cdot\bm{\sigma}$  & $ib$ & $ia$ & $ia\sigma_x+ib\tau_z$ & $iq_0\sigma_0$  & $0$ \\ 
& $\alpha+1=0$ & $\alpha+1=3$   & $\alpha+1=1$ & $\alpha+1=1$ & $\alpha+1=2$ & $\alpha+1=1$ & $\alpha+1=0$  \\ \hline
& \multicolumn{3}{|c|}{$Q\equiv -i\hbar S^\dagger dS/dE=Q^\dagger$} & \multicolumn{4}{|c|}{$Q\equiv -i\hbar \tilde{U}^\dagger (dS/dE) U^\dagger=Q^\dagger=\tilde{Q}$} \\ \hline
$Q_{nm}$ & $q_0$ & $q_0\sigma_0+i\bm{q}\cdot\bm{\sigma}$  & $a+ib$ & $a$ & $a\sigma_0+ib\sigma_y$ & $q_0\sigma_0+i\bm{q}\cdot\bm{\sigma}$ & $a\sigma_0+ib\sigma_y$  \\ \hline
$Q_{nn}$  & $q_0$ & $q_0\sigma_0$  & $a$ & $a$ & $a\sigma_0$  & $q_0\sigma_0$ & $a\sigma_0$ \\ \hline
\end{tabular}
\bigskip

\caption{Characterization of the scattering matrix differential $\delta S$ and of the time-delay matrix $Q$. All coefficients $q_n$, $a,b$ are real, and $\sigma_i$ is a Pauli matrix. The symmetry indices $\beta$ and $\alpha+1$ count, respectively, the number of degrees of freedom of the off-diagonal and diagonal components of the anti-Hermitian matrix $\delta S$. The off-diagonal elements of the Hermitian matrix $Q$ have $\beta$ degrees of freedom, while the diagonal elements have one single degree of freedom in each symmetry class.
}
\label{table_alphabeta}
\end{table*}

\subsection{Broken time-reversal symmetry}

Without the barrier $S_0$ is independent of $Q_0$ and uniformly distributed,
\begin{equation}
P(S_0,Q_0)d\mu(S_0) d\mu(Q_0)=P(Q_0)d\mu(S_0) d\mu(Q_0).\label{PS0Q0start}
\end{equation}
The differential $d\mu$ indicates the Haar measure for the unitary matrix $S_0$ and the Euclidean measure for the Hermitian matrix $Q_0$. The ballistic time-delay matrix distribution is given by \cite{Bro97,Mar14}
\begin{subequations}
\label{PQ0result}
\begin{align}
&P(Q_0^{-1})\propto({\rm Det}'\, Q_0^{-1})^{\alpha+N\beta/2}\nonumber\\
&\qquad\qquad\times\Theta(Q_0)\exp(-\tfrac{1}{2}\beta t_{0}\,{\rm Tr}'\, Q_0^{-1}),\label{PQ0resulta}\\
&\Leftrightarrow P(Q_0)\propto({\rm Det}'\, Q_0)^{-\beta(N-1)-2-\alpha-N\beta/2}\nonumber\\
&\qquad\qquad\times\Theta(Q_0)\exp(-\tfrac{1}{2}\beta t_{0}\,{\rm Tr}'\, Q_0^{-1}).\label{PQ0resultb}
\end{align}
\end{subequations}
Degenerate eigenvalues of $Q_0$ are counted only once in ${\rm Tr}'$ and ${\rm Det}'$. In terms of the degeneracy factor $d$ from Table \ref{table_AZ} this can be written as
\begin{equation}
{\rm Det}'\, Q_0=({\rm Det}\,Q_0)^{1/d},\;\;{\rm Tr}'\,Q_0=\frac{1}{d}\,{\rm Tr}\,Q_0.\label{primeDetTr}
\end{equation}
The channel number $N$ also does not include degeneracies, so the total number of  eigenvalues of $Q_0$ is $d\times N$. The characteristic time $t_0$ differs from the Heisenberg time $t_{\rm H}$ by a numerical coefficient,\footnote{We define $t_{\rm H}=2\pi\hbar/\delta_0$, with $\delta_0$ the mean spacing of \textit{nondegenerate} levels. The ratio $t_0/t_{\rm H}$ then equals the degeneracy of energy levels divided by the degeneracy of delay times \cite{Mar14}. It is unity in all symmetry classes except C and CI, where the delay times have a Kramers degeneracy that the energy levels lack \cite{Bee14}.} 
see Table \ref{table_AZ}.

Insertion of the barrier, with unitary scattering matrix
\begin{equation}
S_{\rm B}=\begin{pmatrix}
r_{\rm B}&t'_{\rm B}\\
t_{\rm B}&r'_{\rm B}
\end{pmatrix},\label{SBdef}
\end{equation}
transforms $S_0$ into
\begin{equation}
\begin{split}
&S=r_{\rm B}+t'_{\rm B}S_0(1- r'_{\rm B}S_0)^{-1}  t_{\rm B}\\
&\Leftrightarrow S_0={t'_{\rm B}}^{-1}S(1-S^{\dagger} r_{\rm B})(1-r_{\rm B}^\dagger S)^{-1}t_{\rm B}^\dagger.
\end{split}
\label{SS0relation}
\end{equation}
Variations of $S$ and $S_0$ are related by \cite{For10}
\begin{equation}
S^\dagger dS={\Sigma} (S_0^{\dagger} dS_0){\Sigma}^\dagger,\;\;{\Sigma}=(1-S^\dagger r_{\rm B})t_{\rm B}^{-1}.\label{dSdS0relation}
\end{equation}

The differentials
\begin{equation}
\delta S=S^\dagger dS,\;\;\delta S_0=S_0^\dagger dS_0,\label{deltaSdef}
\end{equation}
are anti-Hermitian matrices, $\delta S^\dagger=-\delta S$. The number of degrees of freedom of the off-diagonal elements are given by $\beta$ and the number of degrees of freedom of the diagonal elements by $1+\alpha$. As summarized in Table \ref{table_alphabeta}, real matrices (class D) have $\alpha=-1$, $\beta=1$, complex matrices (class A) have $\alpha=0$, $\beta=2$,  and quaternion matrices (class C) have $\alpha=2$, $\beta=4$. These parameters determine the Jacobian \cite{Mat97}
\begin{align}
J_S&=\frac{d\mu(S)}{d\mu(S_0)}=\left|\left|\frac{{\Sigma} \delta S_0{\Sigma}^\dagger}{\delta S_0}\right|\right|\nonumber\\
&=({\rm Det}'\,{\Sigma}{\Sigma}^\dagger)^{(N-1)\beta/2+1+\alpha}.\label{JSdef}
\end{align}

Eq.\ \eqref{dSdS0relation} also implies the relation between the time-delay matrices,
\begin{equation}
Q={\Sigma}Q_0 {\Sigma}^\dagger\Rightarrow dQ={\Sigma}dQ_0 {\Sigma}^\dagger+{\cal O}(dS).\label{QQ0relation}
\end{equation}
The off-diagonal elements of the Hermitian matrix $Q$ have $\beta$ degrees of freedom, the diagonal elements have one single degree of freedom in each symmetry class. The Jacobian is then given by
\begin{align}
J_Q&=\frac{d\mu(Q)}{d\mu(Q_0)}=\left|\left|\frac{{\Sigma}dQ_0 {\Sigma}^\dagger}{dQ_0}\right|\right|\nonumber\\
&=({\rm Det}'\,{\Sigma}{\Sigma}^\dagger)^{(N-1)\beta/2+1}.\label{JQdef}
\end{align}

The joint probability distribution $P(S,Q)$ now follows upon division of $P(S_0,Q_0)$ by the product of Jacobians,
\begin{align}
P(S,Q)&=\frac{P(Q_0)}{J_S J_Q}\nonumber\\
&=P(Q_0)({\rm Det}'\,{\Sigma}{\Sigma}^\dagger)^{-\beta N+\beta-2-\alpha}.\label{PSQJacobian}
\end{align}
Substituting $Q_0={\Sigma}^{-1}Q{{\Sigma}^\dagger}^{-1}$ into Eq.\ \eqref{PQ0result} we thus arrive at the joint distribution
\begin{align}
P(S,Q)&\propto ({\rm Det}'\,{\Sigma}{\Sigma}^\dagger)^{\beta N/2}({\rm Det}'\, Q)^{-3\beta N/2+\beta-2-\alpha}\nonumber\\
&\times\Theta(Q)\exp(-\tfrac{1}{2}\beta t_{0}\,{\rm Tr}'\, {\Sigma}^{\dagger}Q^{-1}{\Sigma}).\label{PSQfinal}
\end{align}
The class-D result \eqref{PS0Q} from the main text follows for $\alpha=-1$, $\beta=1$, $t_0=\tau_{\rm H}$, $d=1$, ${\Sigma}^\dagger={\Sigma}^{\rm T}$.

\subsection{Preserved time-reversal symmetry}

Time-reversal symmetry equates the scattering matrix to its transpose $S^{\rm T}$ in class AI and CI and to its dual $S^{\rm D}$ in class AII and DIII. (The dual of a matrix is $S^{\rm D}=\sigma_y S^{\rm T} \sigma_y$.) We use a unified notation $S=\tilde{S}$, where the tilde indicates the transpose or the dual, whichever is appropriate for that symmetry class. The symmetry $S=\tilde{S}$ allows for the ``square root'' factorization
\begin{equation}
S=\tilde{U}U=\tilde{S},\label{squareroot}
\end{equation}
with unitary $U$.

The time-delay matrix in these symmetry classes is constructed such that it satisfies the same symmetry,
\begin{equation}
Q=-i\hbar \lim_{E\rightarrow 0}\tilde{U}^\dagger\frac{dS}{dE}U^\dagger=\tilde{Q}.\label{QSsymmetry}
\end{equation}
This redefinition of $Q$ differs from Eq.\ \eqref{Qdef} by a unitary transformation, so the delay times are not affected.

The ballistic $Q_0$ and $S_0$ are again independent \cite{Bro97,Mar14}, distributed according to Eqs.\ \eqref{PS0Q0start} and \eqref{PQ0result} with the appropriate values of $\alpha$ and $\beta$ from Table \ref{table_AZ}. These numbers now count the diagonal and off-diagonal degrees of freedom of the symmetrized differential
\begin{equation}
\delta S=\tilde{U}^\dagger dS U^\dagger,\label{deltaSsym} 
\end{equation}
constrained by $\delta S=-\delta S^\dagger$, $\delta\tilde{S}=\delta S$. The matrix elements of $\delta S$ are imaginary in class AI ($\alpha=0$, $\beta=1$), $i$ times a quaternion\footnote{A quaternion has the form $a_0\sigma_0+ia_1 \sigma_x+ia_2\sigma_y+ia_3\sigma_z$, with four real coefficiients $a_n$. The matrix $\sigma_0$ is the $2\times 2$ unit matrix.} in class AII ($\alpha=0$, $\beta=4$), and of the form $a\sigma_x+b\sigma_z$ with $a,b$ imaginary in class CI ($\alpha=1$, $\beta=2$) and $a,b$ real in class DIII ($\alpha=-1$, $\beta=2$).

The elements of the Hermitian matrix $Q$ are real in class AI, quaternion in class AII, and of the form $a\sigma_0+ib\sigma_y$ with $a,b$ real in both classes CI and DIII. The off-diagonal elements of $Q$ have the same number of $\beta$ degrees of freedom as $\delta S$, but the diagonal elements have only a single degree of freedom irrespective of $\alpha$. All of this is summarized in Table \ref{table_alphabeta}.

The symmetrization of the differential modifies the relation \eqref{dSdS0relation}, which now reads
\begin{equation}
\delta S=U{\Sigma}U_0^\dagger \delta S_0 U_0 {\Sigma}^\dagger U^\dagger,\;\; {\Sigma}=(1-U^\dagger\tilde{U}^\dagger r_{\rm B})t_{\rm B}^{-1}.\label{deltaSsymm}
\end{equation}
The relation \eqref{QQ0relation} between $Q$ and $Q_0$ is similarly modified by the symmetrization,
\begin{equation}
Q=U{\Sigma}U_0^\dagger Q_0 U_0 {\Sigma}^\dagger U^\dagger.\label{Qsymm}
\end{equation}
Because the matrices $U$, $U_0$ are unitary, the Jacobians \eqref{JSdef} and \eqref{JQdef} are unchanged,
\begin{align}
J_S&=({\rm Det}'\,{\Sigma}{\Sigma}^\dagger)^{(N-1)\beta/2+1+\alpha},\label{JSdefsym}\\
J_Q&=({\rm Det}'\,{\Sigma}{\Sigma}^\dagger)^{(N-1)\beta/2+1}.\label{JQdefsym}
\end{align}
We thus obtain the joint distribution
\begin{align}
P(S,Q)&\propto ({\rm Det}'\,{\Sigma}{\Sigma}^\dagger)^{\beta N/2}({\rm Det}'\, Q)^{-3\beta N/2+\beta-2-\alpha}\nonumber\\
&\times\Theta(Q)\exp(-\tfrac{1}{2}\beta t_{0}\,{\rm Tr}'\, {\Sigma}^{\dagger}U^\dagger Q^{-1}U{\Sigma}).\label{PSQfinalsym}
\end{align}

\subsection{Poisson kernel}

The marginal distribution of the scattering matrix resulting from the Jacobians \eqref{JSdef} and \eqref{JSdefsym} is
\begin{align}
P(S)&=\int dQ\,P(S,Q)=1/J_S\nonumber\\
&=({\rm Det}'\,{\Sigma}{\Sigma}^\dagger)^{-(N-1)\beta/2-1-\alpha}\nonumber\\
&=\left(\frac{{\rm Det}'(1-r_{\rm B}^\dagger r_{\rm B})}{|{\rm Det}'\,(1-r_{\rm B}^\dagger S)|^2}\right)^{(N-1)\beta/2 +1+\alpha},\label{Poissongeneral}
\end{align}
including the normalization constant. This formula combines the known expressions for the Poisson kernel\footnote{
The name ``Poisson kernel'' applies strictly speaking only to the Wigner-Dyson ensembles, when $r_{\rm B}=\int SP(S)dS$. In the Altland-Zirnbauer ensembles the average scattering matrix differs from $r_{\rm B}$, see Ref.\ \onlinecite{Ber09} and App.\ \ref{app_Saverage}.}
in the Wigner-Dyson ensembles \cite{Bro95} and in the Altland-Zirnbauer ensembles \cite{Ber09}.

The present analysis confirms that Eq.\ \eqref{Poissongeneral} holds without modification in the two symmetry classes D and DIII that support Majorana zero-modes, depending on the sign of the determinant ${\rm Det}\,S=\pm 1$ in class D and the sign of the Pfaffian ${\rm Pf}\,(i\sigma_y S)=\pm 1$ in class DIII \cite{Ful11}. As a check, we can take $N=1$, when $S=\pm 1$ in class D and $S=\pm\sigma_0$ in class DIII. The $\pm$ sign determines the presence or absence of a Majorana bound state (twofold degenerate in class DIII). Since there is only a single element in the ensemble we should have $P(S)=1$, which is indeed what Eq.\ \eqref{Poissongeneral} gives for $N=1$, $\alpha=-1$.

\section{Calculation of the ensemble-averaged scattering matrix}
\label{app_Saverage}

\subsection{Symmetry class D}
\label{sec_rhoavD}

According to Eq.\ \eqref{rhoaverage}, the effect of a tunnel barrier on the average density of states follows directly once we know the average scattering matrix. Simple expressions can be obtained if we assume that the tunnel probabilities are mode-independent, $\Gamma_n=\Gamma$ for $n=1,2,\ldots N$. 

The scattering matrix of the barrier has the polar decomposition
\begin{equation}
S_{\rm B}=\begin{pmatrix}
O_1&0\\
0&O_3
\end{pmatrix}\begin{pmatrix}
\sqrt{1-\Gamma}&\sqrt{\Gamma}\\
\sqrt{\Gamma}&-\sqrt{1-\Gamma}
\end{pmatrix}\begin{pmatrix}
O_2&0\\
0&O_4
\end{pmatrix},\label{SBisotropic}
\end{equation}
with $O_1,O_2,O_3,O_4\in{\rm SO}(N)$. The block structure corresponds to Eqs.\ \eqref{rBdef} and \eqref{SBdef}, in particular, $r_{\rm B}=\sqrt{1-\Gamma}\, O_1O_2$. From Eq.\ \eqref{SS0relation} we obtain the scattering matrix $S$ of the quantum dot,
\begin{equation}
S=O_1\left[\sqrt{1-\Gamma}+\Gamma U(1+\sqrt{1-\Gamma}\,U)^{-1}\right]O_2,\label{StildeS0relation}
\end{equation}
in terms of a matrix $U=O_4 S_0O_3$ that is uniformly distributed in ${\rm O}_\pm(N)$.

Because the average of $U^p$ for any power $p=1,2,\ldots$ is proportional to the identity matrix, we may write the average of $S$ in the form of a power series,
\begin{equation}
\langle S\rangle_\pm=r_{\rm B}\left(1-\frac{\Gamma}{1-\Gamma}\sum_{p=1}^\infty (-1)^{p}(1-\Gamma)^{p/2}\frac{1}{N} \langle{\rm Tr}\,U^p\rangle_\pm \right).\label{SaverageUpaverage}
\end{equation}
If the average of $U$ would be over the entire unitary group, then all terms in the power series would vanish and we would simply have $\langle S\rangle=r_{\rm B}$. But averages over orthogonal matrices do not vanish, in the nontrivial way calculated\footnote{For the record, we note that Eq.\ \eqref{TrUpresult} differs from the formula in Ref.\ \onlinecite{Rai97} by a minus sign ($\pm$ instead of $\mp$).} by Rains \cite{Rai97}: 
\begin{equation}
\langle{\rm Tr}\,U^p\rangle_\pm=\frac{1+(-1)^p}{2}\pm\begin{cases}
(-1)^{N+1}&{\rm if}\;\;p-N=0,2,4,\ldots\\
0&{\rm otherwise}.
\end{cases}\label{TrUpresult}
\end{equation}
We substitute Eq.\ \eqref{TrUpresult} into Eq.\ \eqref{SaverageUpaverage} and sum the geometric series, to arrive at the average scattering matrix
\begin{equation}
\langle S\rangle_\pm=r_{\rm B}\left(1-
\frac{1}{N}\pm\frac{1}{N}(1-\Gamma)^{-1+N/2} \right),\label{Saverageresultapp}
\end{equation}
used in Sec.\ \ref{sec_DOS} to obtain the average density of states in class D.
\\ \\ \\ \\
\subsection{Symmetry class C}
\label{sec_rhoavC}

In a similar way we can derive the average scattering matrix in other symmetry classes. We give the result for class C. The matrix $U$ then varies over the unitary symplectic group ${\rm Sp}(2N)$. Ref.\ \onlinecite{Rai97} gives the required average:
\begin{equation}
\langle{\rm Tr}\,U^p\rangle_{\rm C}=\begin{cases}
-1&{\rm if}\;\;p\leq 2N\;\;\text{and even},\\
0&{\rm otherwise}.
\end{cases}\label{TrUSpresult}
\end{equation}
Eq.\ \eqref{SaverageUpaverage} still holds with the factor $1/N$ replaced by $1/2N$. We thus find the average scattering matrix in class C,
\begin{equation}
\langle S\rangle_{\rm C}=r_{\rm B}\left(1+\frac{1}{2N}[1-(1-\Gamma)^N]\right).\label{SaverageUC}
\end{equation}
For $N=1$ this gives $\langle S\rangle_{\rm C}=(1+\Gamma/2)r_{\rm B}$, in agreement with Ref.\ \onlinecite{Ber09}.

The average density of states in class C follows from Eq.\ \eqref{rhoaveragegeneral}, where we again account for the doubling of the dimensionality $N\mapsto 2N$:
\begin{align}
\langle\rho\rangle_{\rm C}&=\langle\rho\rangle_{\rm ballistic}\,\left(1-\frac{1}{N\Gamma}{\rm Tr}\,r_{\rm B}^\dagger[\langle S\rangle_{\rm C}-r_{\rm B}]\right)\nonumber\\
&=\frac{N}{(N+1)\delta_0}\,\left(1-\frac{1-\Gamma}{N\Gamma}[1-(1-\Gamma)^N]\right).\label{rhoaverageC}
\end{align}
In the second equation we have substituted the ballistic class-C result from Ref.\ \onlinecite{Mar14}. The tunneling limit $\Gamma\rightarrow 0$ gives a vanishing density of states,
\begin{equation}
\langle\rho\rangle_{\rm C}=\frac{N\Gamma}{(N+1)\delta_0}+{\cal O}(\Gamma^2),\label{classCtunnel}
\end{equation}
consistent with the class-C result for a closed system \cite{Boc00,Iva02}.

\end{document}